**Valence-Driven Structural Transformation and Phase Coexistence in Chemically Substituted $EuPd_2Si_2$**

**Masoud Mardani[1], Shivani Sharma[1*], Saul H. Lapidus[2], T. Siegrist[1,3]**
[1]National High Magnetic Field Laboratory, Tallahassee, FL 32310, USA.
[2]Advanced Photon Source at Argonne National Laboratory, Lemont, IL 60439, USA.
[3] Department of Chemical and Biomedical Engineering, FAMU-FSU College of Engineering, Florida State University, Tallahassee FL 32310, USA.
Corresponding author email: phy.shivanisharma@gmail.com

**Abstract:**

$EuPd_2Si_2$ serves as an ideal system to probe the interplay between lattice dynamics and electronic instabilities through its tunable valence transition, responsive to temperature, pressure, and magnetic field. Here, we focus on investigating the origin of the temperature-induced europium valence instability in pristine $EuPd_2Si_2$, alloyed $EuPd_2Si_{1.95}Ge_{0.05}$, and $EuPd_2Si_{1.9}Ge_{0.1}$ and its correlation with structural responses. Compared with the parent compound $EuPd_2Si_2$ which exhibits the europium valence transition at $T_v$ = 143 K, magnetic susceptibility measurements reveal a systematic shift in $T_v$ with Ge substitution, decreasing to 87 K for $EuPd_2Si_{1.95}Ge_{0.05}$. Temperature-dependent synchrotron X-ray diffraction confirms phase coexistence between $Eu^{2+}Pd_2Si_{2-x}Ge_x$ and $Eu^{3+}Pd_2Si_{2-x}Ge_x$ phases across the valence transition for all samples measured (x = 0, 0.05, 0.10). The pronounced increase in the full width at half maximum (FWHM) well above $T_v$ reflects the gradual nature of the transformation and highlights the role of structural strain in stabilizing the coexisting phases, supporting the presence of a phase-segregated ground state, in contrast to a homogeneous intermediate-valence or mixed-valence state. The variation in $T_v$ across different compositions correlates with changes in lattice parameters, highlighting the robust coupling between the structural and electronic degrees of freedom. The phase diagram suggests an inverse linear relationship between the lattice parameter and $T_v$, suggesting a possible predictive framework for estimating transition temperatures in other chemically substituted variants of this family.

## Introduction

The intermetallic compound $EuPd_2Si_2$ (EPS) exhibits a pronounced temperature dependence of the europium (Eu) valence state, where the Eu valence shifts from divalent toward trivalent upon cooling [1–7] while maintaining the $ThCr_2Si_2$ – type structure (Space Group *I4/mmm, No 139*) . This 4f-electron system with the valence transition at $T_v$ ~ 143 K has been heavily studied due to its rich physical properties [8–13]. At room temperature EPS is exhibiting a non-integer valence state close to 2 [1,2,4]. Figure 1a shows the crystal structure of EPS, where selected bond lengths between Eu and Pd/Si atoms are marked, showing their specific values in Ångströms. The effect of the valence transition on these bond lengths is illustrated in Figures 1b and 1c with an overlay of the $Eu^{2+}$ and $Eu^{3+}$- containing 1:2:2 phases by superimposing them in different orientations. The Pd-Si tetrahedral network is shown in 'sea green' color to enhance the local distortions, where the differences between the two lattices have been exaggerated by a factor of eight. The valence transition not only changes the lattice parameters, but certain bond lengths are affected due to volume collapse of the Eu ion, which is visible from figure 1b and 1c.

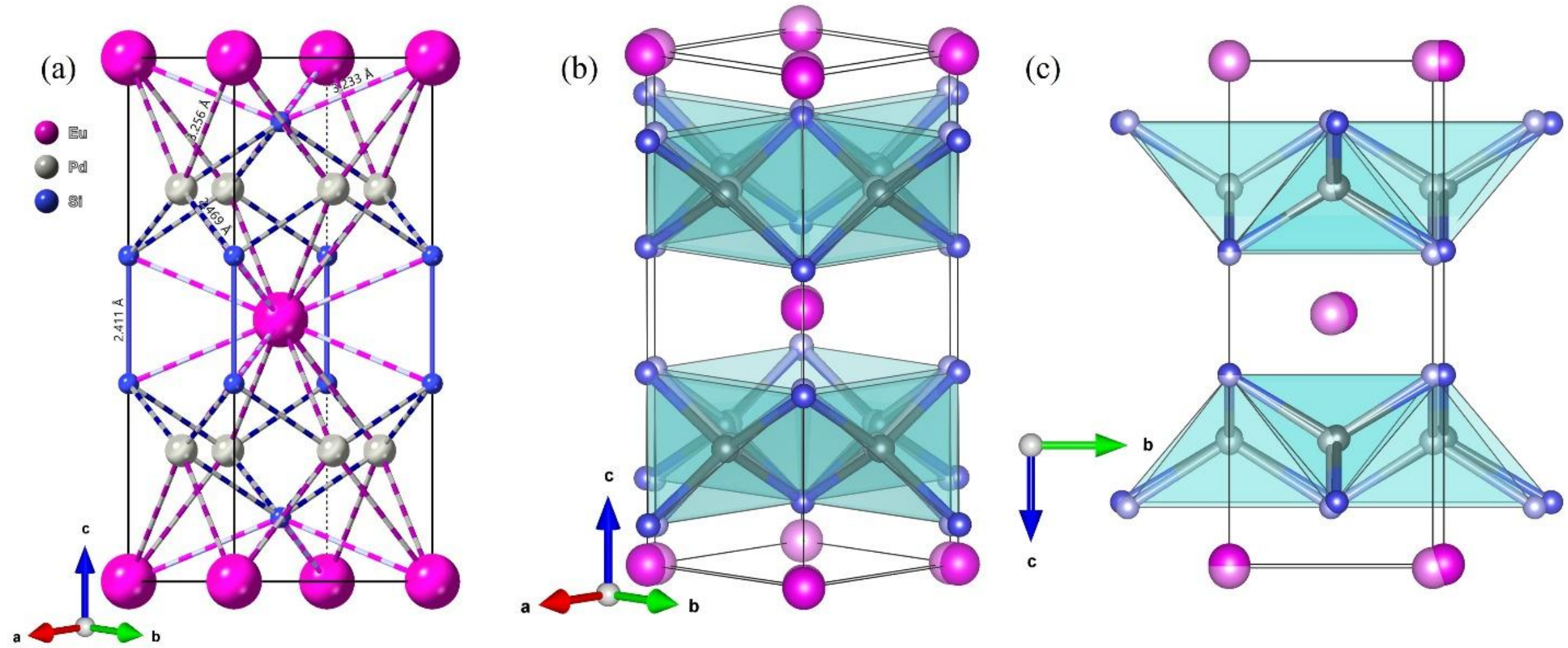

**Figure 1**: (a) Crystal structure of $EuPd_2Si_2$ projected approximately along the [110]-direction, highlighting interatomic distances. The figures in (b) and (c) are created by superimposing the structure of two tetragonal phases belonging to $Eu^{2+}Pd_2Si_2$ and $Eu^{3+}Pd_2Si_2$ to visualize the changes in various bonds in different orientation. The dark color pink, grey and blue atoms are Eu, Pd and Si atoms respectively, belonging to $Eu^{2+}Pd_2Si_2$ whereas the light color atoms belong to $Eu^{3+}Pd_2Si_2$ phase.

Given the substantial ionic radius difference between $Eu^{2+}$ and $Eu^{3+}$, large lattice distortions are expected to accompany the valence state transition as discussed above. Moreover, the magnetic response of divalent $Eu^{2+}$ (J = 7/2) contrasts sharply with the non-magnetic $Eu^{3+}$ (J = 0), highlighting the strong interplay between the valence state transition, magnetic susceptibility, electronic transport and structural properties in these compounds. Interestingly, several compounds in this family, such as $EuPd_{2-x}Pt_xSi_2$, $EuRh_2Si_2$, $EuCo_2Ge_2$ and $EuNi_2Ge_{2-x}Si_x$ exhibit valence transition under external applied pressure and/or external magnetic field [3,4,14–18]. Among these phases, EPS is the most extensively studied, owing to its tunable valence transition, which can be controlled by temperature, magnetic field, and chemical substitution [3,18,19]. Regardless of the tuning parameter, the valence transitions are expected to be accompanied by substantial changes in the lattice dimension and changes in the Si atom position, underscoring the key role of the crystal structure. Since the ionic radii of $Eu^{2+}$ is significantly larger than for $Eu^{3+}$, a reduction in unit cell volume [9], and the delocalization of a 4*f* electron into the 5*d* band of Eu is observed upon cooling.

In EPS, substituting elements with different atomic radii at the Pd or Si sites produce chemical pressure effects similar to external pressure. This induced chemical pressure is expected to affect the electronic properties, which are further associated with modifications in the magnetic and structural responses. Previously, Wada et al. reported a lowering of $T_v$ upon 15% Rh substitution at the Pd site whereas no significant change has been observed in a 10% Ru substituted compound [19]. Similarly, compounds such as (Eu:RE)$Pd_2Si_2$, (Eu:RE)$Cu_2Si_2$, and (Eu:RE)$NiSi_2$ (RE = La, Y, and Yb) with low Eu concentrations demonstrate the importance of the nearest neighboring environment and unit cell volume in determining the valence state of Eu [20]. Furthermore, in $Eu(Pd_{1-x}Pt_x)_2Si_2$, a systematic decrease in $T_v$ with increasing Pt concentration (x) has been reported [21] while Ge substitution on the Si site leads to a more substantial reduction in $T_v$ [3]. High-field magnetization studies on $Eu(Pd_{1-x}Pt_x)Si_2$ further demonstrated a universal scaling behavior between the magnetic field and temperature, suggesting a single energy scale governs the valence transition across different tuning parameters [18]. Notably, Ge substitution for Si in EPS also leads to a decrease in the magnetic field required to drive the transition, providing direct evidence of the coupling between lattice and electronic degree of freedom [22].

A closely related phenomenology has long been recognized in samarium sulfide (SmS), which represents one of the earliest and most extensively studied examples of a pressure-induced valence transition in rare-earth systems [23–28]. SmS undergoes a first-order isostructural transition from a divalent, semiconducting state to a metallic state under applied pressure, accompanied by a large volume collapse and pronounced lattice anomalies. Importantly, this transition exhibits phase coexistence and hysteresis over a finite pressure range, underscoring the crucial role of strain accommodation and electron–lattice coupling in governing the valence instability. The SmS system therefore provides a paradigmatic framework for understanding how competing elastic, electronic, and thermal effects can give rise to heterogeneous states or domain formation during a valence transition.

EPS represents a chemically tunable intermetallic analogue of SmS, where similar physics can be explored as a function of temperature, magnetic field, and chemical pressure rather than hydrostatic pressure alone. Structurally, the valence transition in EPS strongly affects the *a*-axis lattice parameter, with an overall change of 0.04 Å, a relative change of the order of 1%. Such a large change indicates that for a fluctuating europium valence, the associated atomic motion due to the volume collapse and the motion due to thermal activation are in competition. The resulting anharmonic motion leads to a large thermal expansion coefficient in the phase transition region. However, the transition extends over a sizeable temperature range, indicating that the transition may have a substantial phase coexistence. If a transition from $Eu^{2+}$ to $Eu^{3+}$ is nucleated, the expected strain in the vicinity will drop significantly, delaying the valence transition of nearby europium atoms. We therefore expect that nucleation of the $Eu^{3+}$ valence state will result in a domain structure, with residual $Eu^{2+}$ retained well below the transition temperature. As a result, the system is expected to become heterogeneous, with both $Eu^{2+}$ and $Eu^{3+}$ coexisting below $T_v$, with a persistent paramagnetic signature below $T_v$ In contrast, if the valence transition is viewed as a continuous transition where the Eu 4*f* states hybridize with the Eu 5*d* states, intermediate valence would result. In this case, the Eu valence would adopt continuous values between 2 and 3, with fluctuations present, and a uniform change in the lattice dimensions.

In this work, we employ temperature-dependent high resolution synchrotron x-ray diffraction as well as laboratory-based systems and magnetic susceptibility to investigate the evolution of the structure as a function of temperature. The thermal expansion probes the anharmonicity of the bonding potential, and via the thermal expansion coefficient, it can be considered as a way of exerting an effective pressure. Since thermal expansion will approach zero at low temperature, where the atomic motion is predominantly harmonic, the effective pressure change is expected to diminish at low temperatures. We further use isoelectronic substitutions to exert chemical pressure on the europium atoms in the structure by substituting Ge on the Si site in nominal concentration. Germanium substitution will decrease the chemical pressure by expanding the lattice. The structural results of two compounds $EuPd_2Si_{2-x}Ge_x$ (x = 0.05, 0.10) are compared with the parent EPS. While previous investigations using X-ray absorption spectroscopy (XAS), magnetic susceptibility, and Mössbauer spectroscopy have provided critical insights into the electronic and magnetic aspects of valence fluctuations [6,13,29–33], these techniques are predominantly sensitive to electronic states. In contrast, high-resolution XRD offers a direct window into the structural changes, including variations in structural parameters such as lattice constants and bond lengths. Therefore, our work provides a structural perspective, allowing for a more comprehensive understanding of the correlation between electronic behavior and the underlying lattice dynamics driving valence instability in Eu-based intermetallics.

## Experimental

Polycrystalline samples of $EuPd_2Si_2$ (EPS), $EuPd_2Si_{1.95}Ge_{0.05}$ (EPS-Ge2.5%) and $EuPd_2Si_{1.9}Ge_{0.10}$ (EPS-Ge5%) were prepared by arc melting [9,10,22]. The as prepared ingots were sealed under argon in quartz tubes for annealing at 875 ºC for 8 days. The phase purity of the samples was checked by powder X-ray diffraction (XRD) using a Seifert/Eigenmann X-ray diffractometer with Cu-Kα radiation and an energy dispersive detector. The bulk magnetization measurements were carried out in a Quantum Design MPMS SQUID magnetometer as a function of temperature and magnetic field. Heat capacity measurements have been performed on EPS using Quantum Design Physical Property Measurement System. Low temperature XRD studies have been conducted at the 11-BM beamline at the advanced photon source (APS, Argonne National Laboratory) using an X-ray wavelength of $\lambda$ = 0.4582 Å in a temperature range of 87 K to 298 K. Structural properties below 87 K for EPS and EPS-Ge2.5% were investigated by powder XRD measurements with a Huber 670 Guinier Camera equipped with a helium closed cycle cryostat and a Cu-$K\alpha_1$ lab source in the temperature range of to 10 K to 300 K. Rietveld refinements of the XRD data were performed using the JANA program suite [34].

## Results and Discussion:

Figure 2 shows the Rietveld refinement of room temperature XRD patterns of EPS, EPS-Ge2.5% and EPS-Ge5% in the tetragonal space group symmetry *I*4/*mmm*. Small amounts of $Pd_2Si$ are found in all the samples as a secondary phase and is included in the refinement. The presence of $Pd_2Si$ does not interfere with the main structural features. In Figure 2, the tetragonal phase at room temperature is referred to as the high temperature (HT) tetragonal phase as $Eu^{2+}Pd_2Si_2$. In the crystal structure (see Figure 1), Eu atoms occupy the 2a (0,0,0) Wyckoff position, Pd atoms occupy the 4d (1/2,0,1/4) position, and Si/Ge atoms are located at the 4e (0,0,z) position. The refined structural parameters, along with anisotropic thermal parameters for all the compositions are given in Table 1. The values are consistent with the single-crystal data of Ge-substituted $EuPd_2Si_{1-x}Ge_x$ [22]. As expected, the lattice parameters *a* and *c* increase with Ge substitution at the Si site. Selected bond lengths for all compositions are listed in Table S1 in the supplemental information [35].

**Table1:** Structural parameters for different compositions, determined from synchrotron powder data. The anisotropic atomic displacement parameters ($U_{ii}$) are included.

| Parameters | EPS | EPS-Ge2.5% | EPS-Ge5% |
|---|---|---|---|
| | Lattice parameters (Å) | | |
| a (Å) | 4.24204(3) | 4.24841(3) | 4.25341(6) |
| c(Å) | 9.87959(9) | 9.89522(8) | 9.8839(1) |
| | Eu: 2a (0,0,0) | | |
| $U_{11}$ (Å$^2$) | 0.0087(1) | 0.0063(2) | 0.0171(2) |
| $U_{33}$ (Å$^2$) | 0.0164(3) | 0.0200(5) | 0.0298(5) |
| | Pd: 4d (0.5,0,0.25) | | |
| $U_{11}$ (Å$^2$) | 0.0068(1) | 0.0077(2) | 0.0155(2) |
| $U_{33}$ (Å$^2$) | 0.0208(3) | 0.0148(4) | 0.0301(6) |
| | Si,Ge: 4s (0,0,z) | | |
| z | 0.3779(1) | 0.3780(2) | 0.3744(2) |
| $U_{11}$ (Å$^2$) | 0.0057(4) | 0.0047(6) | 0.0051(7) |
| $U_{33}$ (Å$^2$) | 0.0087(9) | 0.0150(13) | 0.0219(14) |
| | Goodness of fit parameters | | |
| GOF | 1.54 | 2.13 | 1.69 |
| $R_p$ | 8.54 | 10.18 | 9.37 |

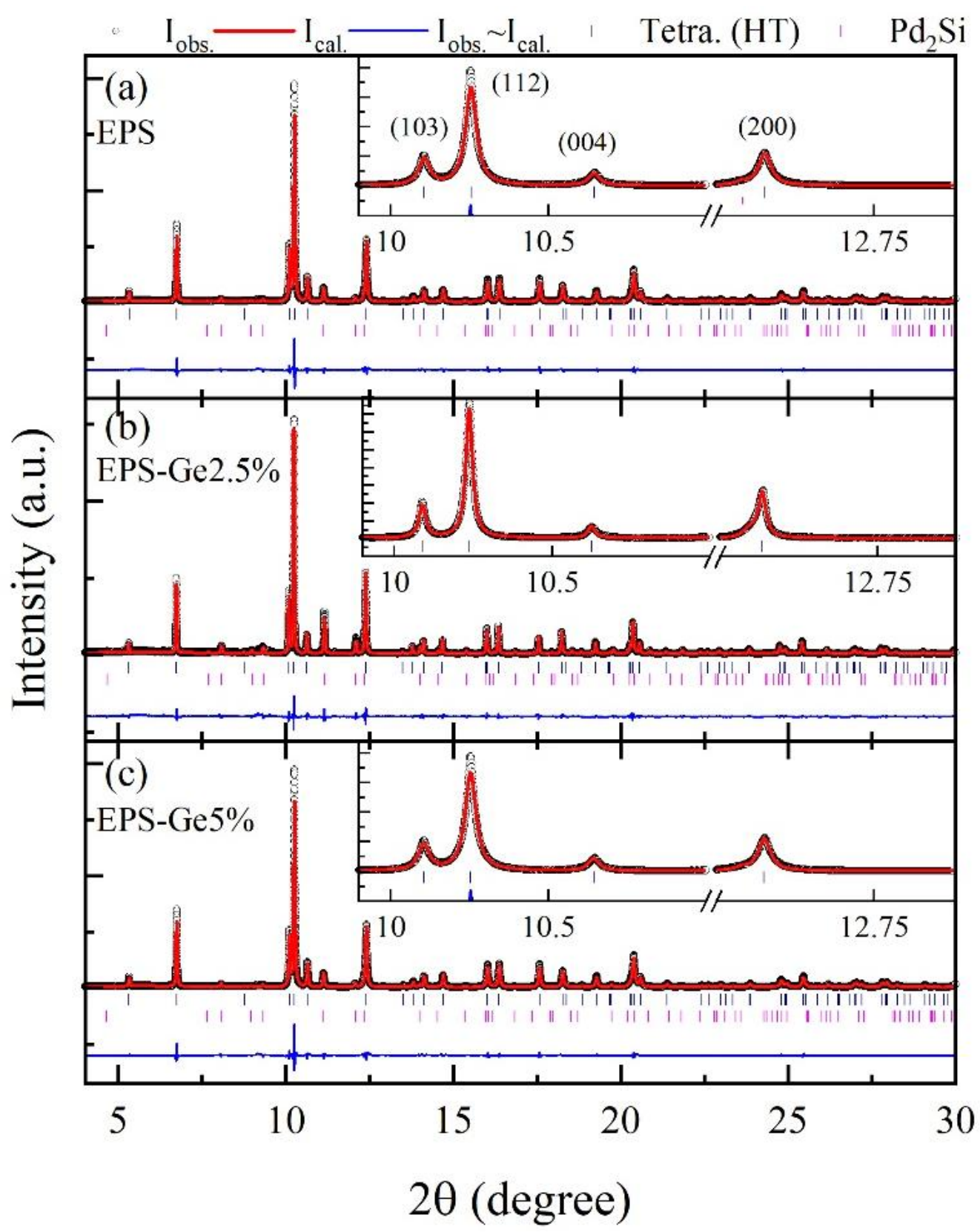

**Figure 2**: Rietveld refined synchrotron X-ray patterns taken at room temperature of (a) EPS, (b) EPS-Ge2.5% and (c) EPS-Ge5% using a two-phase model ($Eu^{2+}Pd_2Si_2$+$Pd_2Si$). The inset in each figure presents the zoom view of four selected peaks (103), (112), (004) and (200), indicating the good quality of the overall fit.

Figure 3 shows the magnetic susceptibility $\chi(T)$ of EPS and EPS-Ge2.5%. For both samples, the magnetic susceptibility increases upon cooling, followed by a broad peak with a maximum at $T_v$. For EPS, $T_v$ = 143 K and for EPS-Ge2.5% $Tv$ = 87 K, consistent with the valence transition temperature reported previously [18,22]. For EPS, upon further cooling, the magnetic susceptibility increases again continuously down to 2 K, which is shown in the inset of figure 3a. The magnetization curve at 2 K closely follows the expected Brillouin function for $Eu^{2+}$ ($4f^7$, S = 7/2, L = 0, J = 7/2), with an observed saturation moment of 2.17 $\mu_B$ per formula unit. This is nearly one-third of the theoretical expected value (7 $\mu_B$) for fully ordered $Eu^{2+}$. This reduced moment indicates either an $Eu^{2+}Pd_2Si_2/Eu^{3+}Pd_2Si_2$ mixed-valence ground state or a heterogeneous material where approximately one-third of the Eu ions remain in the $Eu^{2+}$ configuration and contribute to the magnetic moment, while the two-thirds are $Eu^{3+}$ ($4f^6$, J = 0) ions in the non-magnetic $Eu^{3+}Pd_2Si_2$ phase. This one third concentration of $Eu^{2+}Pd_2Si_2$ phase is consistent with structural results which will be discussed in the subsequent sections. If the valence transition were complete at $T_v$, no substantial Curie-Weiss behavior would be expected below $T_v$ due to the J=0 state of $Eu^{3+}$. Therefore, divalent Eu remains present in the sample at the lowest temperatures, but no antiferromagnetic order is observed.

For both the samples, the isothermal magnetization at 300 K exhibits a paramagnetic response, whereas at 2 K the curve closely follows the Brillouin function expected for $Eu^{2+}$ ($4f^7$, S = 7/2, L = 0, J = 7/2) only for EPS.

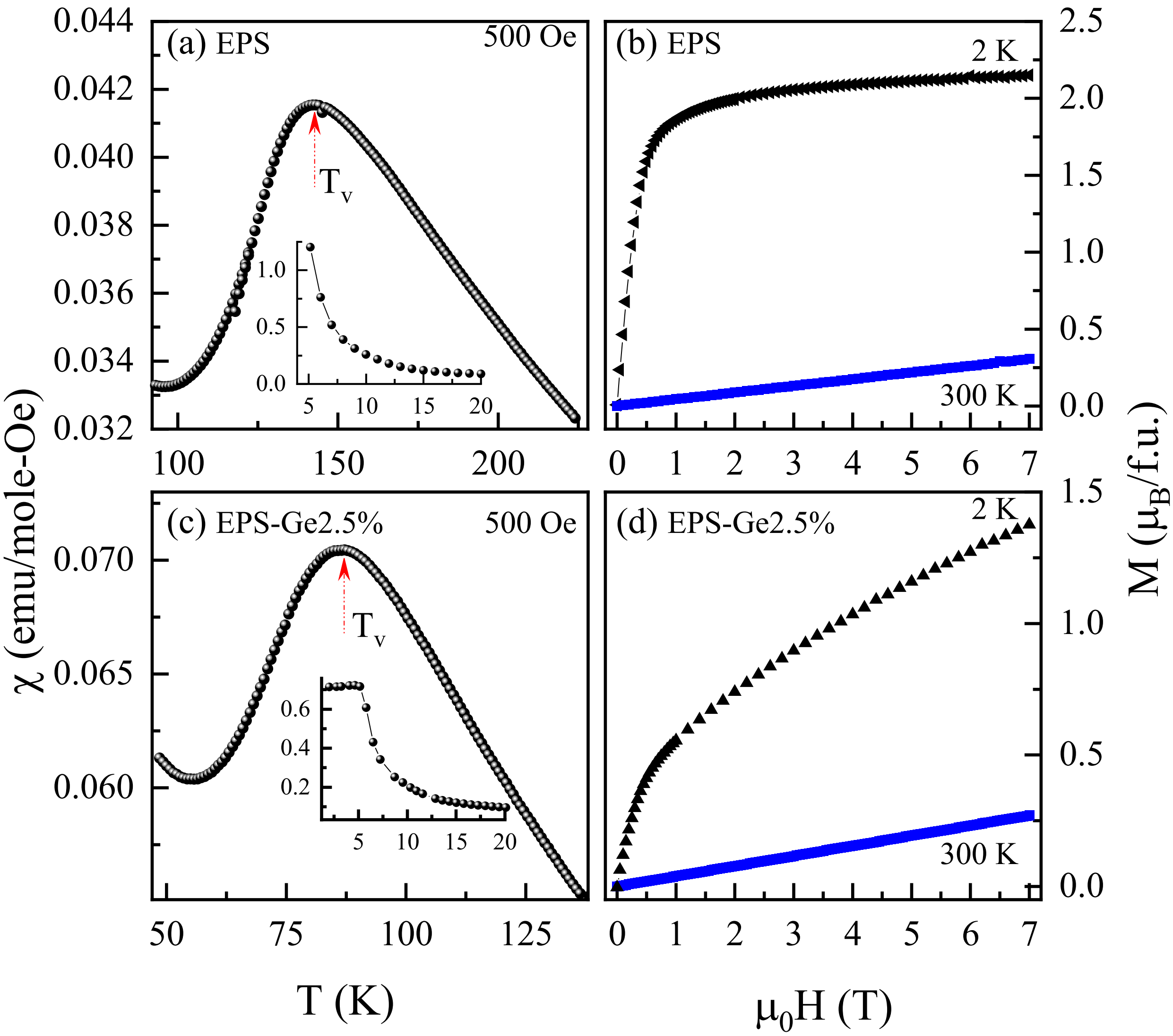

**Figure 3:** Magnetic susceptibility of (a,b) EPS and (c,d) EPS-Ge2.5% as a function of temperature (χ-vs-T) and magnetic field (*M*-vs-*H*). The inset in (a) and (c) presents the low temperature $\chi$-vs-$T$ data for respective samples.

The significant shift in $T_v$ with a small 5% Ge substitution on the Si site highlights the strong sensitivity of the transition to lattice dimensions. For EPS-Ge2.5%, $T_v$ = 87 K, and at lower temperatures, the magnetic susceptibility shows a similar behavior as EPS. In contrast to EPS, saturation near 5 K is observed. This feature may have two possible origins: since a larger fraction of divalent Eu persists down to 2 K, it is likely that these $Eu^{2+}$ moments undergo antiferromagnetic ordering near 5 K. Related systems with stable $Eu^{2+}$ valence, such as $EuPd_2Ge_2$ show antiferromagnetic order below 17 K [22]. Alternatively, the behavior could arise from a minor magnetic impurity not detectable in XRD. However, this is unlikely in view of the magnitude of the observed moment. Furthermore, the non-saturating magnetization curve at 2 K is consistent with a magnetically ordered state (Figure 3d).

The heat capacity of EPS exhibits a pronounced anomaly below $T_v$, centered at 125 K, providing clear thermodynamic evidence for the characteristic Eu valence transition (Figure 4). The large, narrow heat-capacity peak and the abrupt change in heat capacity across the transition are characteristic of a first-order thermodynamic transition which will be further confirmed based on the structural properties. In $EuPd_2Si_2$, this anomaly is associated with the well-established temperature-induced Eu valence transition. The pronounced peak reflects the substantial entropy involved in this transformation and is consistent with the first-order character of the valence transition.

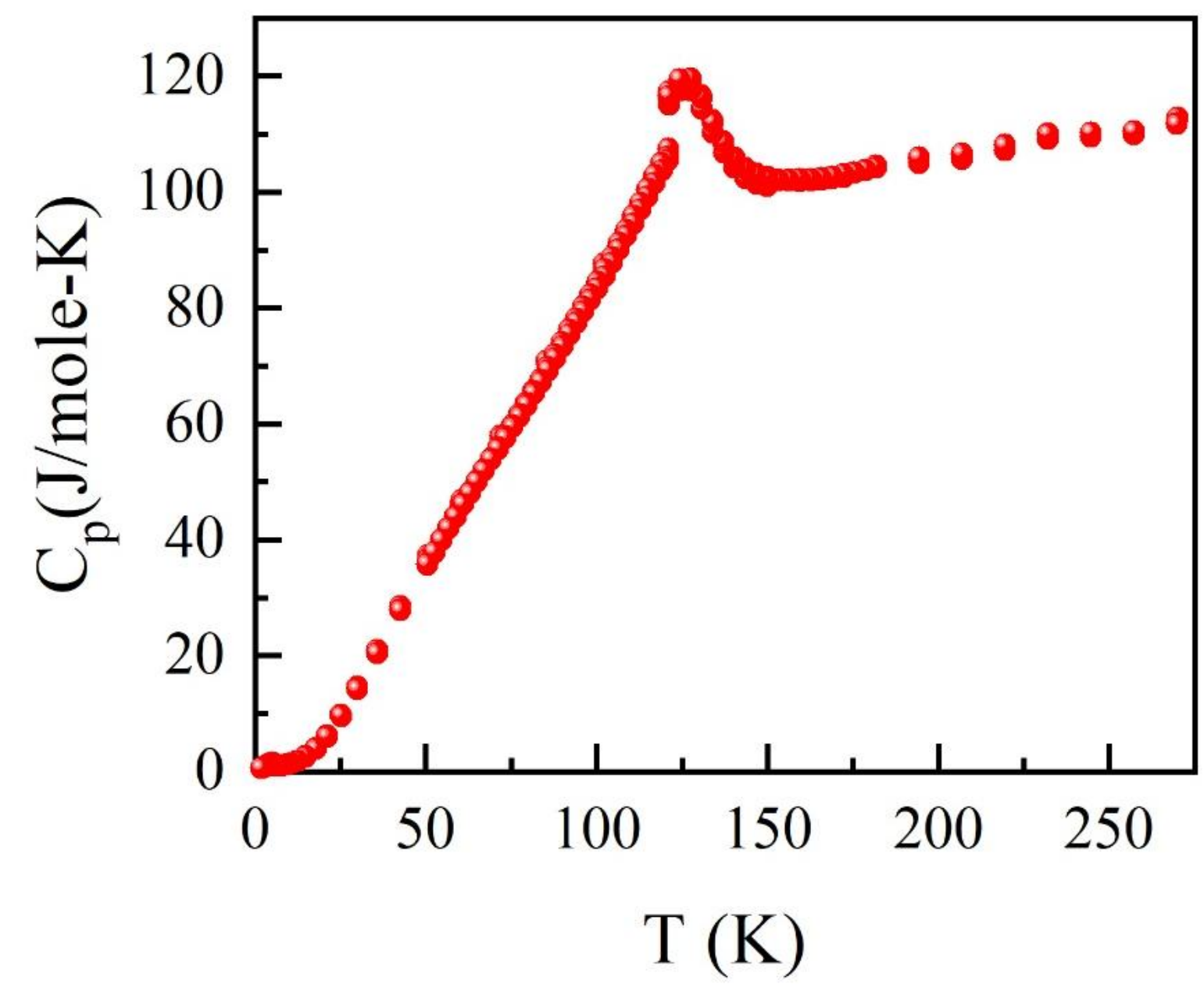

Figure 4: Heat capacity of EPS as a function of temperature. The latent heat around $T_v$ is released/taken up over an extended temperature range.

If a homogenous mixed or intermediate valent state is present in EPS with a random distribution of $Eu^{2+}$/$Eu^{3+}$ atoms, a single unit cell parameter set (a,c) is expected. In contrast, if two separate phases with $Eu^{2+}$ and $Eu^{3+}$ form, two different unit cell sets of (a,c) will be present below $T_v$ in a phase separated system. To test this hypothesis, synchrotron powder XRD patterns were recorded at T = 298, 200, 150, 120 and 87 K. Figure 5a shows a stack of the (200) and (004) reflections as a function of temperature, while the complete XRD patterns are provided in the supplementary information as figure S1 [35]. The (004) peak shows negligible shifts at the valence transition and only slightly broadens upon cooling of the sample. In contrast, the (200) peak gradually broadens upon cooling the sample to 150 K, and below $T_v$, it splits into two distinct peaks. The broadening of the reflections is indicative of a gradual onset of the valence change above $T_v$. Furthermore, the (200) reflection is clearly split below 100 K. The reflection at higher 2θ corresponds to the $Eu^{3+}Pd_2Si_2$ with a smaller lattice parameter *a*, while the reflection at the lower 2θ is associated with $Eu^{2+}Pd_2Si_2$ with a larger *a* parameter. At 87 K, $Eu^{3+}Pd_2Si_2$ is clearly dominant. In the absence of high-quality synchrotron data with a large signal-to noise ratio and superior resolution, the two distinct peaks may not be fully resolved, and only peak broadening and an average shifted peak position may be observed, as reported in previous studies [1,22,36].

This phase coexistence observed during the valence state transition has often been overlooked in the past due to experimental limitations.

Figure 5b shows the temperature dependence of the full width at half maximum (FWHM) of the (200) peak, associated with the high temperature $Eu^{2+}Pd_2Si_2$ phase. The FWHM reflects the changes in structural coherence and internal strain, which are often tied to the transition and usually serve as an order parameter. For temperatures where two distinct peaks were observed, the FWHM was determined by fitting the data with two reflection profiles and the FWHM of the peak at lower 2θ is plotted in figure 5b, corresponding to the $Eu^{2+}Pd_2Si_2$ phase, which is the main phase at room temperature. The FWHM starts increasing significantly at a temperature far above $T_v$ which indicates the onset of enhanced strain leading up to the transition. However, the most pronounced change in FWHM of both reflections occurs below the transition temperature which is observed as an increase below 120 K as shown in figure 5b, indicating enhanced strain.

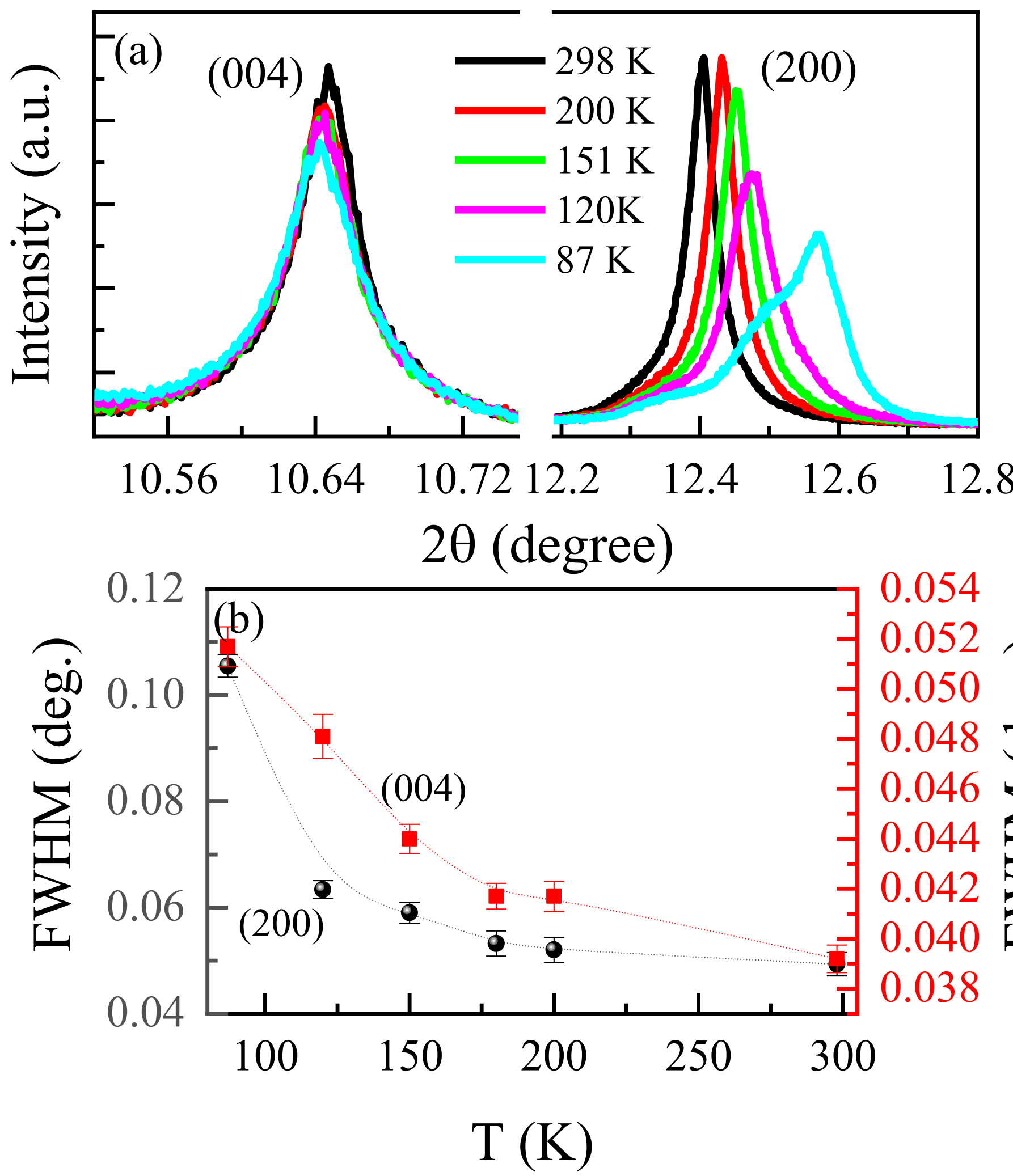

**Figure 5:** (a) Stacked reflection profiles of the (004) and (200) reflections at selected temperatures. (b) Temperature dependence of the full width at half maximum (FWHM), highlighting a gradual increase well above below $T_v$, with a sharp rise below 120 K. For temperatures where two peaks

are resolved, the FWHM of the lower 2θ component ($Eu^{2+}$-rich phase) is plotted. The dashed line in (b) serves as a guide to the eye. The intrinsic FWHM of the instrument is of the order of 0.04°.

Rietveld refinement of the temperature-dependent XRD data was carried out to quantify the evolution of the crystal structures and coexistence of phases. The refined pattern at 87 K is shown in figure 6a. This refinement employed a three-phase model that includes both $Eu^{2+}Pd_2Si_2$, $Eu^{3+}Pd_2Si_2$, plus the minor $Pd_2Si$ impurity phase. For clarity and consistency, we refer to $Eu^{2+}Pd_2Si_2$ as the "high-temperature tetragonal" (Tetra. HT) phase and $Eu^{3+}Pd_2Si_2$ as the "low-temperature tetragonal" (Tetra. LT) phase, reflecting the different lattice parameters. The high quality of the Rietveld fit at 87 K can be clearly seen in the inset of figure 6a, highlighting the (112) and (200) reflections—both of which show peak splittings consistent with the presence of two coexisting tetragonal phases. In contrast, the (004) reflection shows a minimal shift during the transition, indicating that the structural evolution is primarily driven by changes in the $a$-axis parameter.

The temperature dependence of the refined lattice parameter $a$ is given in figure 6b. The variation in the $c$ parameter as a function of temperature is negligible and shown in figure S2 of the supplemental information, along with the derived $c/a$ ratio [35]. As expected, the lattice parameter $a_{HT}$ decreases with decreasing temperature, reflecting conventional thermal contraction down to $T_v$. Below $T_v$, two phases with different lattice parameters ($a_{HT}$ and $a_{LT}$) coexist which can be clearly seen for T = 100 K and 87 K in figure 6b, confirming the phase separation into $Eu^{2+}$ and $Eu^{3+}$-rich domains. This strongly supports a scenario in which the phase coexistence across the valence transition results in distinct structural motifs at low temperature, rather than persisting as a homogeneous mixed-valence state. Since the refined lattice parameters from the laboratory Guinier system are very well matched above and below the transition with the data derived from the synchrotron measurement, it is reasonable to use the lab source data to obtain the relative thermal expansion coefficient $\alpha, \left(= \frac{1}{a}\frac{da}{dT}\right)$above the transition. To calculate $\alpha_{HT}$ of the HT phase, 175 and 290 K data points were used, whereas for $\alpha_{LT}$ for the LT phase, 20 and 60 K data points were used. The values of $\alpha_{HT}$ and $\alpha_{LT}$ are $2.8904*10^{-5}$ and $3.0309*10^{-5}$ per Kelvin respectively. Since the Debye temperature is of the order of 215 K, the thermal expansion should approach zero below 50 K [37]. Since the thermal expansion coefficient is related to the heat capacity through the Grüneisen relation,

$$\alpha = \frac{\gamma\, C_V}{B\, V} \qquad (1)$$

where $\gamma$ is the Grüneisen parameter, $B$ is the bulk modulus, and $V$ is the volume, the thermal expansion coefficient also decreases rapidly at low temperatures and approaches zero as $T \to 0$. Since $T/\Theta_D \approx 0.05$ at 10 K, the phonon contribution to the thermal expansion is expected to be strongly suppressed in this temperature range for a valence-stable system. Therefore, any measurable low-temperature change in the lattice parameter is unlikely to arise solely from conventional lattice vibrations and may instead reflect the additional contribution from the continued evolution of the Eu valence state.

A rough estimation of thermally induced pressure needed to drive this valence state transition can be made using the thermal expansion coefficient. Barrera et al. have derived the mathematical expression where thermal expansion can be expressed as the elastic response to thermally induced stress [25]. If the isothermal compressibility is $\chi_T$, then the volumetric thermal expansion coefficient $\alpha$ can be expressed by

$$\alpha = \chi_T \left(\frac{\partial P}{\partial T}\right)_V \quad (2)$$

The value of $\chi_T$ can be estimated from the bulk modulus B ($\chi_T = 1/B$). For $EuPd_2Si_2$, the reported values of bulk modulus lie in the range 70–135 GPa. Taking a representative average value of B ~ 100 GPa, $\alpha = 3 * 10^{-5}\ K^{-1}$, the thermally induced pressure change is estimated to be of order ~ 300 MPa for a temperature variation of 100 K. This estimate provides a useful scale for comparing the effects of thermal effect and internal pressure on the valence transition.

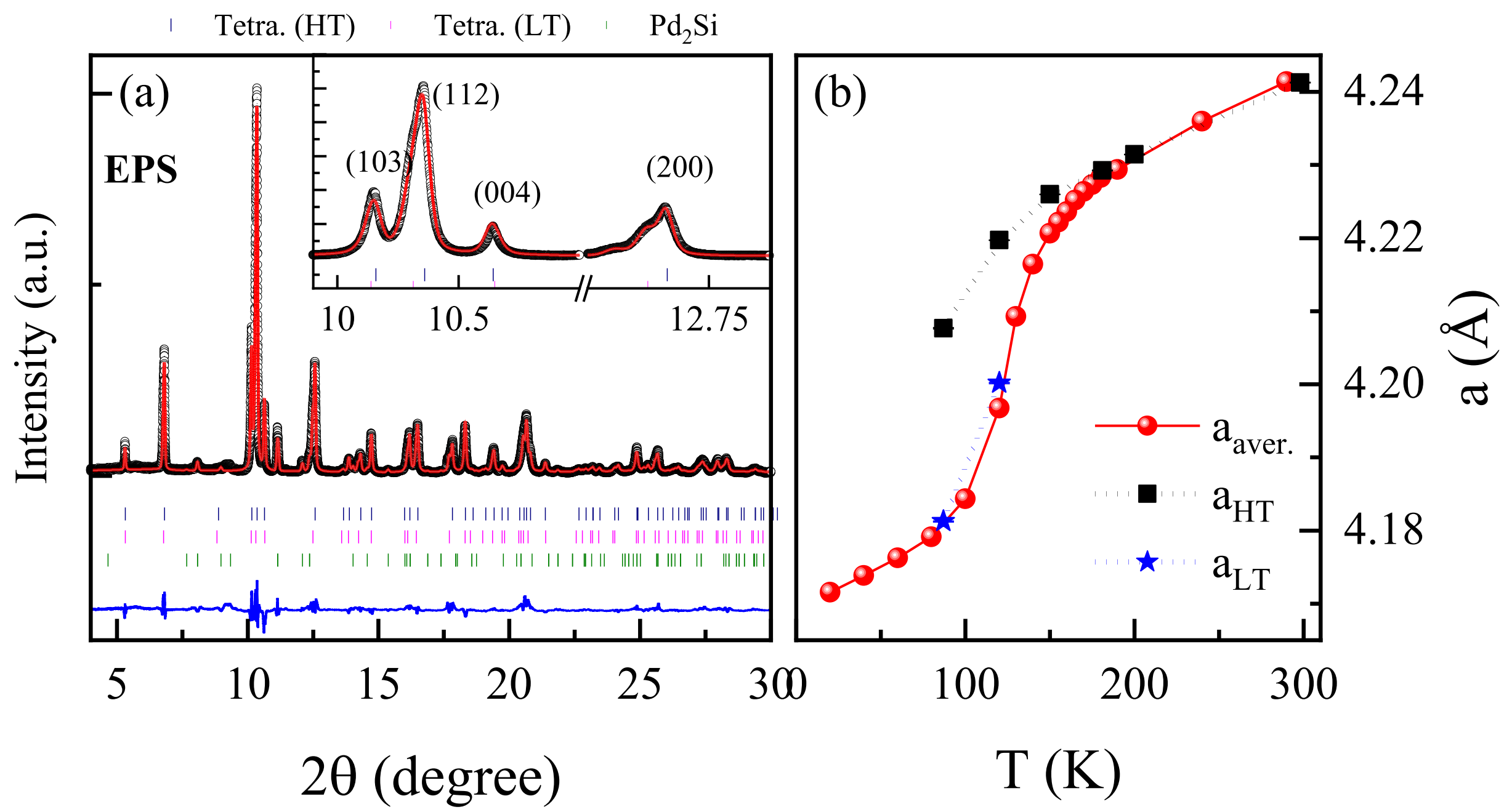

**Figure 6**: (a) Rietveld refinement of XRD data at 87 K using a three-phase model: $Eu^{2+}Pd_2Si_2$ (Tetra. HT), $Eu^{3+}Pd_2Si^2$(Tetra. LT), and $Pd_2Si$ impurity. The inset shows peak splitting in the (112) and (200) reflections, consistent with coexistence of two tetragonal phases, while the (004) peak remains nearly unchanged. (b) Temperature dependence of the lattice parameter *a*. Below $T_v$, the divergence of *a* values confirms separation into $Eu^{2+}$- and $Eu^{3+}$-rich domains, supporting spatially resolved phase coexistence over an intermediate-valence state.

Previous studies on $EuPd_2Si_2$ (EPS) reported a significant reduction in the lattice parameter *a* across the valence transition without considering the coexistence of two Eu phases near $T_v$ [38]. In contrast, the high-resolution synchrotron XRD data clearly resolve the splitting of reflections into two phases with distinct lattice parameters, revealing that the apparent lattice contraction originates from the evolution of their relative phase fractions. To demonstrate this, we also refined our laboratory XRD data, collected with a typical FWHM resolution of ~0.08°. The waterfall plot of these XRD patterns is given the supplemental information as figure S3 [35]. The resulting average lattice parameter, $a_{aver.}$ shown in Fig. 6b exhibits the same continuous decrease upon cooling reported previously [13]. While $a_{aver.}$ agrees with $a_{HT}$ above 200 K and $a_{LT}$ below $T_v$, it deviates in the intermediate temperature range, masking the underlying phase coexistence. The broad coexistence regime and peak broadening observed even above $T_v$ further suggest that strain acts as a precursor to the transition and may explain the residual paramagnetic response below $T_v$ arising from the persistence of $Eu^{2+}$-rich regions. Such a distinct anomaly is characteristic of a first-order phase transition and provides strong thermodynamic evidence supporting its first-order nature. Furthermore, lattice parameter changes below 50 K in EPS should not be observable due to harmonic atomic motion. The observed changes shown in Figure 6b are a strong indication that the valence transition has very wide temperature range and the ration of $Eu^{3+}/Eu^{2+}$ increases below $T_v$.

The strain-driven phase coexistence observed in $EuPd_2Si_2$ highlights the crucial role of lattice perturbations in governing the valence transition. The Ge substitution provides an effective means of tuning the lattice through chemical pressure. Owing to the larger atomic size of Ge, even modest substitution results in significant lattice expansion, with the lattice parameter *a* increasing by ~0.15% for 2.5% Ge and ~0.26% for 5% Ge substitution. This compares well with the change in the lattice parameters *a* during the valence transition in EPS, which is of the order of 1%. Refined lattice parameters for different compositions at room temperature can be found in Table 1. By expanding the unit cell volume, it is expected that $T_v$ will shift to lower temperatures since the thermal contraction (expansion) will reach the critical pressure at a lower temperature. Indeed, for EPS-Ge2.5%, $T_v$ is significantly reduced to $T_v$ = 87 K as shown in figure 3(c,d), a change of 56 K.

In Figure 7, the peak profiles of (200) reflections and its FWHM are plotted together for EPS-Ge2.5%and EPS-Ge5%. The FWHM values in figure 7(c,d) correspond to the reflection associated with the high-temperature $Eu^{2+}Pd_2Si_2$ phase. Similar to EPS, a distinct splitting of the (200) peak is evident at 100 and 87 K for EPS-Ge2.5% whereas no such splitting is observed down to 87 K for EPS-Ge5%. Nevertheless, the FWHM plotted in the right-hand panel for each composition clearly increases with decreasing temperature, suggesting that the structural instability is developing in EPS-Ge5% below 87 K. The gradual increase in FWHM upon cooling indicates a progressive trend consistently observed across all compositions.

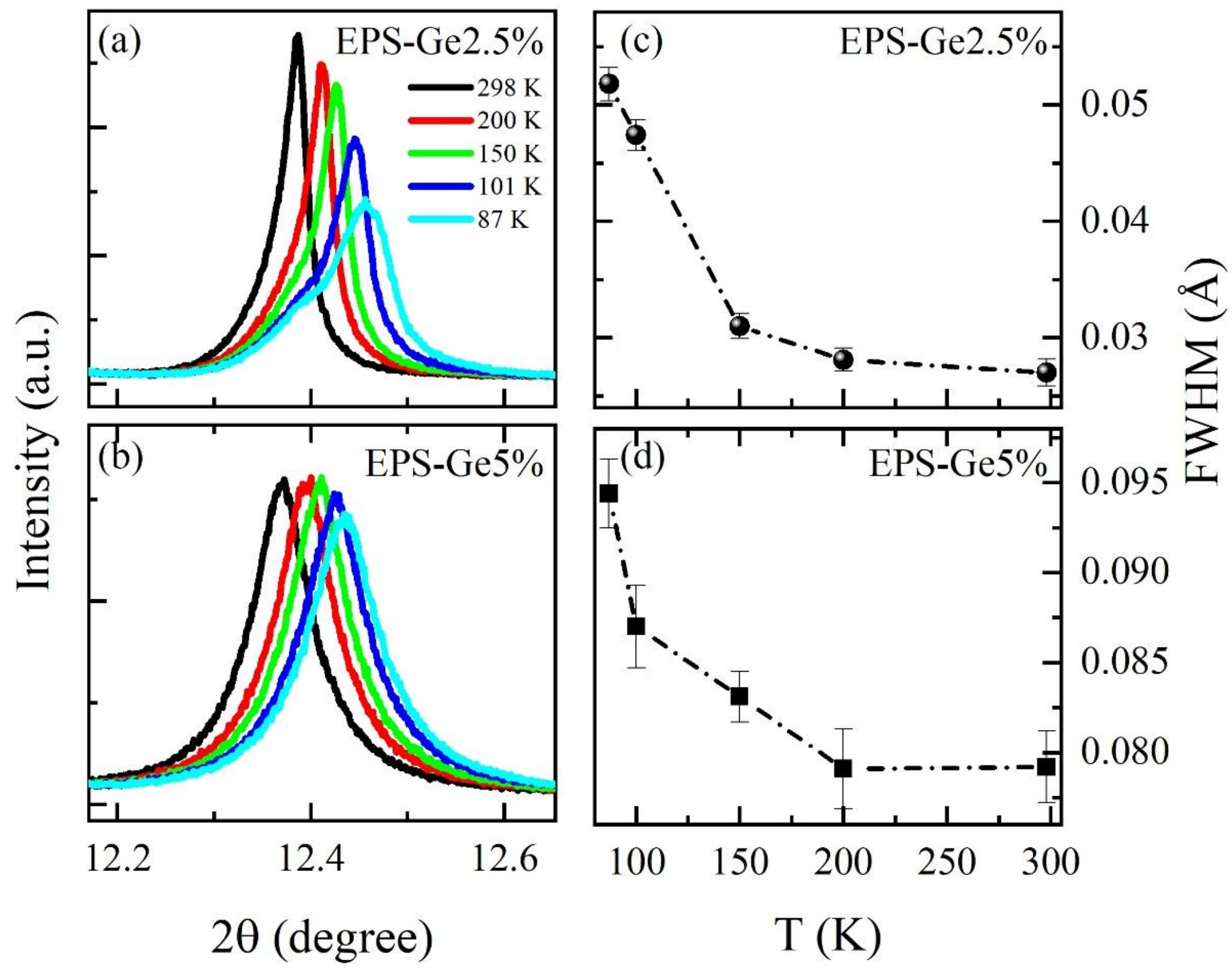

**Figure 7**: Temperature dependence of (a,b) (200) peak profile and (c,d) full width at half maximum (FWHM) for different compositions. The dashed line in the right-side panel is a guide to the eye.

The temperature dependence of the lattice parameters *a* and relative phase fraction (PF) of HT $Eu^{2+}Pd_2Si_2$ and LT $Eu^{3+}Pd_2Si_2$ phases is shown in Figure 8(a-b). Filled symbols in Figure 8b represent the phase fraction of the HT tetra phase, while open symbols indicate the phase fraction of the LT tetra. phase. The refined XRD patterns of EPS-Ge2.5% and EPS-Ge5% at 87 K are given in the supplemental information as Fig. S4 [35]. Upon cooling, the lattice parameter *a* decreases with temperature for all samples down to ~100 K. Below this temperature, it splits into two distinct sets of values, corresponding to the coexistence of two tetragonal phases both for EPS and EPS-Ge2.5%. A single lattice parameter *a* has been used to refine the data down to 87 K. The lattice parameter *c* again exhibits minimal variation across $T_v$ for Ge substituted samples, similar to EPS as shown in Fig. S2 of supplemental material [35].

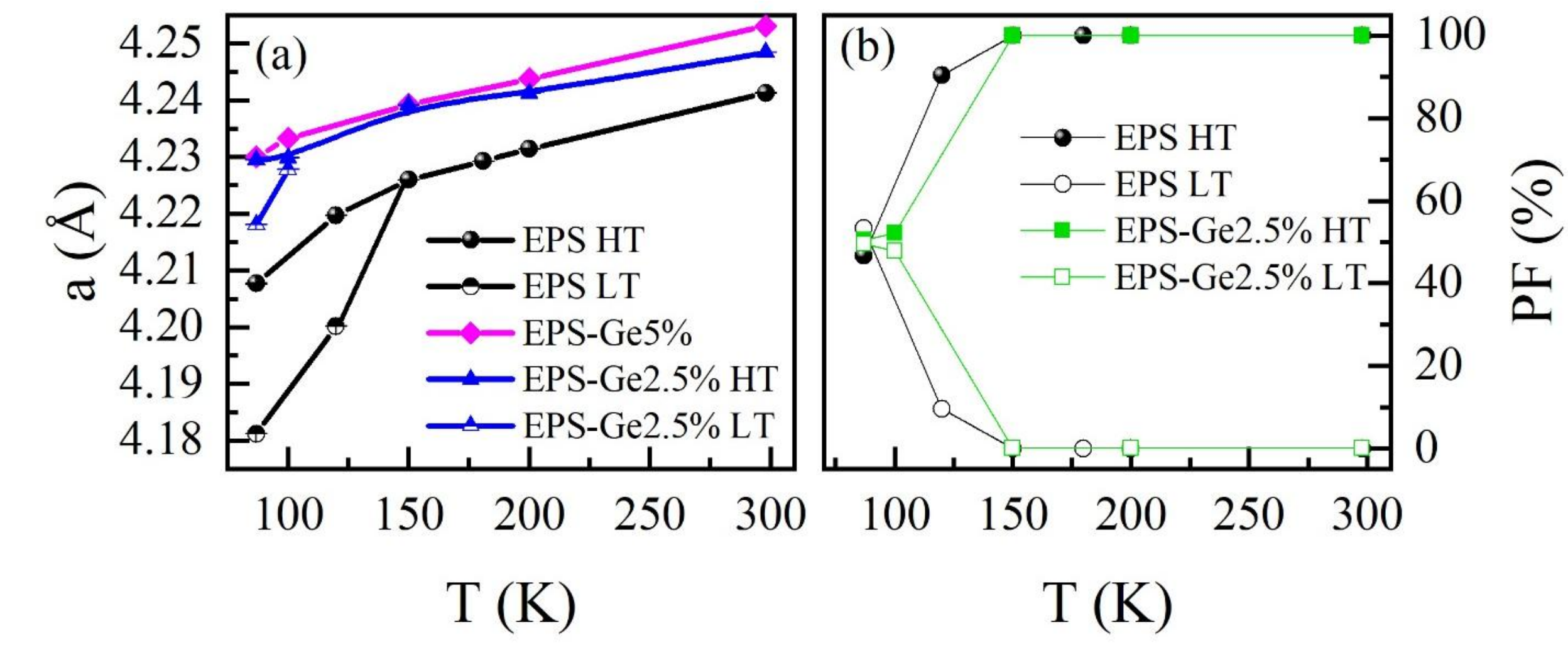

**Figure 8:** (a) Rietveld-refined XRD peak profiles for EPS-Ge2.5%, and EPS-Ge5%, showing selected reflections fitted with high accuracy. (b) Temperature-dependent phase fractions of the high-temperature (HT) and low-temperature (LT) tetragonal phases for the respective compositions. (c,d) Temperature evolution of the lattice parameters *a* and *c* for different compositions.

While the expected valence transition temperature for EPS-Ge5% is below 87 K, the lowest accessible temperature for the synchrotron measurements, we further investigated the growth of the $Eu^{3+}Pd_2(Si,Ge)_2$ phase in EPS-Ge5% using the laboratory Guinier system to a temperature of 10 K. Figure S5 shows a stack of XRD profiles for selected reflections over the temperature range of 290 to 10 K for EPS-Ge5%. As the temperature decreases, the progressive broadening of the (200) peak is observed, and below 60 K, the peak distinctly splits into two components. This splitting is even more pronounced than that seen in EPS, likely due to enhanced chemical pressure arising from site disorder in EPS-Ge5%. Details of the analysis using a two-phase model is discussed in the supplemental information [35]. Additionally, lattice parameter changes are observed well below 60 K, indicating that the ratio of $Eu^{3+}/Eu^{2+}$ is still increasing as the temperature is lowered. The associated volume collapse is therefore driving observed “thermal expansion / contraction”.

To further elucidate the structural response to the valence transition, selected bond lengths and the average *z* coordinate are plotted as a function of temperature in Fig. 9. In the crystal structure, the only refinable atomic coordinate is the *z*-parameter of the Si/Ge atom. During the two-phase refinement involving $Eu^{2+}Pd_2Si_2$ and $Eu^{3+}Pd_2Si_2$, the *z*-parameter exhibited minor changes around the average value. Therefore, only the values corresponding to the average structure are plotted. A clear trend is observed across all samples: the *z*-value increases as the temperature decreases, indicating that the structure can accommodate the strain produced by the thermal contraction. This variation in *z* directly influences several bond lengths, which can be understood through their underlying relationships [35]. A distinct decrease in Eu–Pd, Eu–Si, and Pd–Pd bond lengths is observed below $T_v$ for EPS, showing the structural responses to the valence transition. In the Ge-substituted compounds, a similar decreasing trend is evident for EPS-Ge2.5%, while for EPS-Ge5%, the accessible temperature range of the synchrotron data does not extend low enough to capture the anticipated bond length contraction, as the refinement of the laboratory data did not give values with sufficient precision.

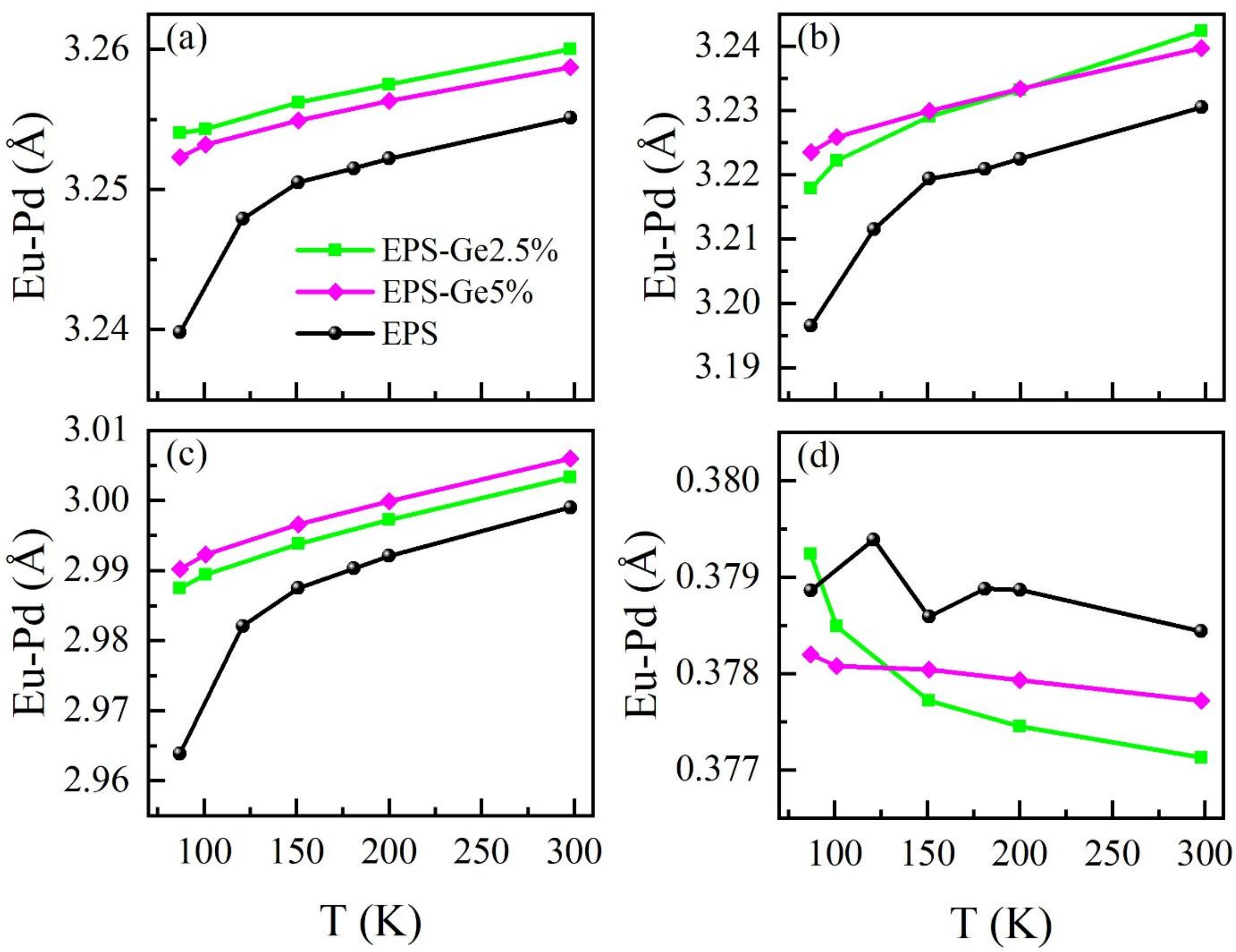

**Figure 9**: (a-c) Selected bond lengths for different compositions, and (d) the z parameter of Si atoms as a function of temperature.

Based on the structural and magnetization data, diagram has been constructed that relates the lattice parameter a to $T_v$ and is shown in Fig. 10. The diagram shows the correlation between the $T_v$ and the lattice parameter $a$. For clarity, the right vertical axis shows $1/a$, whereas $T_v$ is plotted on the left vertical axis. An overall trend is observed in which larger lattice parameters are associated with lower transition temperatures. This behavior suggests that $T_v$ is sensitive to chemical pressure produced by isovalent substitutions. In particular, the 5% Ge-substituted sample, which exhibits the largest lattice parameter among the compositions studied, also shows the lowest transition temperature, consistent with this trend.

To further examine this correlation, $T_v$ has been plotted as a function of lattice parameter $a$ and fitted using a linear approximation. The fit yields a slope of $-7.355(5) \times 10^3$ K/Å (inset of Fig. 10). The negative slope reflects the observed inverse tendency between $T_v$ and $a$, although its quantitative significance should be interpreted with caution given the limited number of compositions. Within the limited composition range studied here, this relationship is an empirical trend rather than a general scaling behavior. Using this empirical fit, an estimated $T_v$ of approximately 48 K is obtained for a lattice parameter of $a$ = 4.255 Å. Further studies across a broader compositional space will be required to establish whether this trend extends beyond the present system or is specific to the present set of substituted compounds.

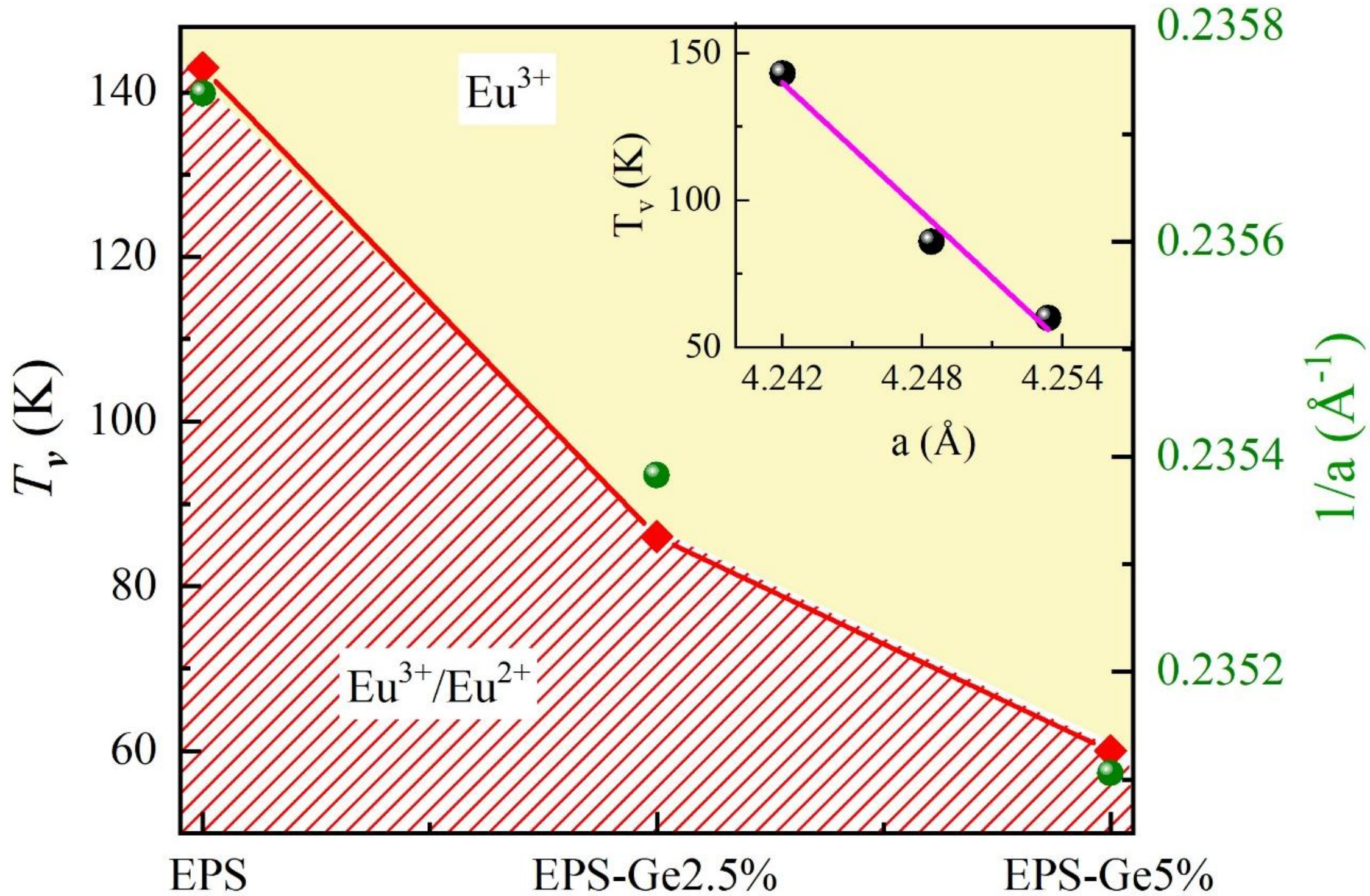

**Figure 10**: Phase diagram of $EuPd_2Si_{2-x}Ge_x$. The plot shows the valence state transition temperature $T_v$ (left vertical axis, in red) and the inverse lattice parameter $1/a$ (right vertical axis, in green) as a function of chemical substitution. The hatched region indicates the phase coexistence regime. The inset displays the variation of $T_v$ with the lattice parameter $a$. The pink line represents a linear fit to the data."

## Conclusion

Temperature-dependent synchrotron XRD analysis confirms a clear structural phase coexistence between $Eu^{2+}Pd_2Si_2$ and $Eu^{3+}Pd_2Si_2$ phases across the valence transition, evidenced by peak splitting and broadening of the (200) reflection with an extended transition range in temperature. The significant increase in FWHM well above $T_v$ indicates the progressive nature of the transition and the role of structural strain, emphasizing phase coexistence rather than valence fluctuations and random mixed valence. The anisotropic lattice evolution, predominantly along the *a*-axis, and the clear separation of lattice parameters support a phase-separated ground state over a homogeneous mixed-valence phase. Comparison with lower-resolution laboratory XRD data shows that average lattice parameter analysis can mask this structural complexity, underscoring the necessity of high-resolution measurements to resolve phase coexistence and associated strain effects over a broad temperature range.

Negative chemical pressure by isovalent substitution of Si by the larger Ge atom suppresses the valence transition temperature, where a significant drop in $T_v$ is observed for EPS-Ge2.5% ($T_v$ = 87 K), confirming the strong coupling between lattice, electronic, and magnetic degrees of freedom. Minor changes in the Si/Ge *z*-parameter with decreasing temperature impact bond lengths and bond angles, underscoring the strong correlation between lattice distortions and the structural mechanisms driving the valence transition. An empirical scaling of $T_v$ with the lattice parameter *a* was established within the investigated compositional range, providing a useful relationship for estimating the transition temperature in related substituted compounds.

Our findings provide insight into the microscopic nature of the valence transition in $EuPd_2Si_2$ and its derivatives, emphasizing the critical role of lattice strain and phase coexistence in governing the complex interplay of valence, structure, and magnetism in this family of compounds. A Debye equation of state indicates that below 50 K, no thermal expansion should occur. In contrast, the changes in the lattice parameter *a* at these temperatures is clearly due to the valence instability, producing the “anomalous” thermal expansion/contraction at low temperatures in these systems.

**Acknowledgement**

A portion of this work was performed at the National High Magnetic Field Laboratory, which is supported by National Science Foundation Cooperative Agreement No. DMR-2128556 and the State of Florida. M.M and T.S. acknowledges funding from the National Science Foundation under Grant No. DMR-2219906. S.S. acknowledges support from the Center for Rare Earths, Critical Minerals, and Industrial Byproducts at the National High Magnetic Field Laboratory, Florida State University.